\documentclass[12pt]{iopart}
\usepackage{amsfonts}
\usepackage{graphicx}
\newcommand{\bnabla}{\mbox{\boldmath $\nabla$}}
\begin{document}

\title[Induced vacuum current and magnetic field]{Induced vacuum current and magnetic field
in \\ the background of a cosmic string}

\author{Yu A Sitenko$^1$ and N D Vlasii$^2$}
\address{$^{1}$ Bogolyubov Institute for Theoretical Physics,
National Academy of Sciences, \\ 14-b Metrologichna Str., Kyiv,
03680, Ukraine}
\address{$^{2}$ Physics Department, National Taras Shevchenko University of
Kyiv, \\ 2 Academician Glushkov Ave., Kyiv, 03022, Ukraine}
\ead{yusitenko@bitp.kiev.ua}

\begin{abstract}
Vacuum polarization effects in the cosmic string background are
considered. We find that a current is induced in the vacuum of the
quantized massive scalar field and that the current circulates
around the string which is generalized to a $(d-2)$-brane in locally
flat $(d+1)$-dimensional space-time. As a consequence of the Maxwell
equation, a magnetic field strength is also induced in the vacuum
and is directed along the cosmic string. The dependence of the
current and the field strength on the string flux and tension is
comprehensively analyzed. Both the current and the field strength
are holomorphic functions of the space dimension, decreasing
exponentially with the distance from the string. In the case of
$d=3$ we show that, due to the vacuum polarization, the cosmic
string is enclosed in a tube of the magnetic flux lines if the mass
of the quantized field is less than the inverse of the transverse
size of the string core.
\end{abstract}

\submitto{\CQG}

\section{Introduction}
Cosmic strings are topological defects which are formed as a result
of phase transitions with spontaneous breakdown of symmetries in
early universe, see, e.g., reviews in \cite{Hi,Vi}. Starting with a
random tangle, the cosmic string network evolves into two distinct
sets: the stable one which consists of several long, approximately
straight strings spanning the horizon volume and the unstable one
which consists of a variety of string loops decaying by
gravitational radiation. A straight infinitely long cosmic string in
its rest frame is characterized by stres-energy tensor with only two
nonvanishing components $T_{00}=-T_{33}$, where the 3-rd coordinate
axis is chosen to be directed along the string. The appropriate
global characteristics of the string is tension (or linear density
of mass)
\begin{equation}
\mu=\int\limits_{\rm core}d^2x\sqrt{g}\, T_{00},\label{eq1}
\end{equation}
where the integration is over the transverse section of the core of
the string, and in what follows we use units $c=\hbar=1$ and the
metric conventions of \cite{Mis}. According to general relativity,
stress-energy tensor is a source of gravity, and, consequently, the
space-time region corresponding to the string core possesses
positive scalar curvature $R=16\pi GT_{00}$ ($G$ is the
gravitational constant), since energy density $T_{00}$ is positive
inside this region. In the case of a cosmic string associated with
spontaneous breakdown of a continuous symmetry, tension (1) is
related to mass $m_{\rm H}$ of the Higgs scalar field, $\mu \sim
m_{\rm H}^2$, and, in addition, the string is also characterized by
gauge field with strength $B^3=\varepsilon^{ii'3}\partial_{i}V_{i'}$
($\bf V$ is the gauge field vector potential and $V_0=0$) which is
directed along the string and is nonvanishing inside its core
\cite{Nie}. The appropriate global characteristics is flux of this
gauge field strength
\begin{equation}
\Phi=\int\limits_{\rm core}{\rm d}^2x\sqrt{g}\, B^3;\label{eq2}
\end{equation}
note that the transverse size of the string core is of order of
$m^{-1}_{\rm H}$. The space-time metric outside the string core is
defined by squared length element
\begin{equation}
\fl {\rm d}s^2=-{\rm d}t^2+(1-4G\mu)^{-1}{\rm d}{\stackrel{\sim}{r}}^2+
(1-4G\mu)\stackrel{\sim}{r}^2{\rm d}\varphi^2+{\rm d}x^2=
-{\rm d}t^2+{\rm d}r^2+r^2{\rm d}\stackrel{\sim}{\varphi}^2+{\rm d}x^2,\label{eq3}
\end{equation}
where
\begin{equation}
\stackrel{\sim}{r}=r\sqrt{1-4G\mu}, \quad 0\leq\varphi<2\pi,\quad
0\leq\stackrel{\sim}{\varphi}<2\pi (1-4G\mu).\label{eq4}
\end{equation}
A surface which is transverse to the string is isometric to the
surface of a cone with a deficit angle equal to $8\pi G\mu$. Such
space-times were known a long time ago (M.~Fierz, unpublished, see
footnote in \cite{We}) and were studied in detail in \cite{Ma1,Ma2}.
In the present context, as cosmological objects appearing after
phase transitions and under the name of cosmic strings, they were
introduced in seminal works of Kibble \cite{Ki1,Ki2} and Vilenkin
\cite{Vil1,Vil2} (see also \cite{Zel,Gar}). The interest to this
subject is recently revived owing to the finding that many
supergravity (superstring) models of inflation predict the
production of cosmic strings at the end of inflationary phase
\cite{Sar,Jean,Pol}. Cosmic strings may also arize in other
approaches such as string gas cosmology \cite{Bra} and dark matter
strings \cite{Vach}. The nonvanishing of string tension (1) leads to
various cosmological consequences and, among them, to a very
distinctive gravitational lensing effect \cite{Ma2,Vil1} (see
\cite{Sa} for current discussion) and to a specific form of
discontinuities in the temperature of the cosmic microwave
background radiation \cite{Steb}. Namely the latter effect imposes
the upper bound on the value of tension:
$\mu\mathrel{\mathop{<}\limits_{\sim}} 10^{-7}G^{-1}$ (see
\cite{Bat}).

As already mentioned, flux $\Phi$ is nonvanishing for the strings
corresponding to spontaneous breakdown of local (continuous)
symmetries. If tension vanishes ($\mu=0$), then such a string
becomes similar to a magnetic string, i.e. a tube of the magnetic
flux lines in flat space. If the tube is impenetrable for quantum
matter, then quantum effects outside the tube may depend on flux
$\Phi$ periodically with period $2\pi e^{-1}$ ($e$ is the coupling
constant \,--\, charge of the matter particle). This is known as the
Bohm-Aharonov effect \cite{Aha}, which has no analogue in classical
physics, since the classical motion of charged particles cannot be
affected by the magnetic flux from the impenetrable for the
particles region. The natural question is, how the nonvanishing
string tension ($\mu\neq0$) influences quantum effects of the
string. Thus, the subject of cosmic strings, in addition to
tantalizing phenomenological applications, acquires a certain
conceptual importance.

Quantum-mechanical motion of charged particles in the background of
a cosmic string was considered in \cite{Sou,Si2,Si5,SiV}. In the
present paper we are interested in the effects of the
second-quantized theory and consider the vacuum polarization which
is induced by a cosmic string in quantum matter. The vacuum
energy-momentum tensor in the cosmic string background was
considered in \cite{Fro,Dow,Gui1,Gui2}, and our concern will be in
the vacuum current in this background. We shall assume that the
effects of the string core structure are negligible since the
transverse size of the core is estimated to be of several orders
less than the size of a proton and can be neglected; thus our
consideration will be restricted to idealized (infinitely thin)
cosmic strings and some of the transverse size effects will be
discussed afterwards. On the other hand, in view of the significance
of high-dimensional space-times in various aspects, we shall deal
with a generalization of a cosmic string to spaces of arbitrary
dimensions. Quantum matter will be represented by charged massive
scalar field.

\section{Generalization to space-times of arbitrary dimensions}

The generalization of an idealized cosmic string to a
$d$-dimensional space is given by a $(d-2)$-brane in the conical
$(d+1$)-dimensional space-time with squared length element
\begin{equation}
    {\rm d}s^2=-{\rm d}t^2+{\rm d}r^2+(1-4G\mu)^2r^2{\rm d}\varphi^2+{\rm d}{\bf x}_{d-2}^2,\label{eq5}
\end{equation}
where $r$ and $\varphi$ are polar coordinates of conical surface,
and ${\bf x}_{d-2}$ are Cartesian coordinates of flat
$(d-2)$-dimensional space; thus the spatial curvature is
nonvanishing and singular in the $(d-2)$-brane (i.e., point in the
$d=2$ case, line in the $d=3$ case, plane in the $d=4$ case, and
$(d-2)$-hypersurface in the $d>4$ case). The gauge field strength
(bundle curvature) is generalized to be directed along the brane,
i.e. bundle connection ($V_r,\,V_\varphi,\,{\bf V}_{d-2}$) is taken
in the form
\begin{equation}
  V_r=0,\,\,V_\varphi=\Phi/2\pi,\,\,{\bf V}_{d-2}=0,\label{eq6}
\end{equation}
and, appropriately, bundle curvature $B^{i_3\ldots
i_d}=\varepsilon^{i_1\ldots i_d}\partial_{i_1}V_{i_2}$ is
nonvanishing and singular in the brane:
\begin{equation}
  B^{3\ldots d}(r,\,\varphi,\,{\bf x}_{d-2})=\Phi\frac{\delta(r)}{(1-4G\mu)r}\Delta(\varphi).\label{eq7}
\end{equation}
Here $\Phi$ is the total flux of the bundle curvature and is given
by (2) with $B^{3\ldots d}$ substituted for $B^3$;
$\Delta(\varphi)=(2\pi)^{-1}\sum\limits_{n\in\mathbb{Z}}{\rm
e}^{{\rm i}n\varphi}$ is the delta-function for compact (angular)
variable, and $\mathbb{Z}$ is the set of integer numbers.

The operator of a second-quantized charged scalar field is presented
in the form (see, e.g., \cite{Itz})
\begin{equation}
   \Psi(t,\,{\bf x})=\sum\!\!\!\!\!\!\!\int\limits_{\lambda}\frac{1}{\sqrt{2E_\lambda}}
  \left[{\rm e}^{-{\rm i}E_\lambda t}\psi_\lambda({\bf x})a_\lambda+
  {\rm e}^{{\rm i}E_\lambda t}\psi_{-\lambda}({\bf x})b^\dag_\lambda\right].\label{eq8}
\end{equation}
Here $a_\lambda^\dag$ and $a_\lambda$ ($b_\lambda^\dag$ and
$b_\lambda$) are the scalar particle (antiparticle) creation and
destruction operators satisfying commutation relations
\begin{equation}
\left[a_{\lambda},a^\dagger_{\lambda'}\right]_-=
\left[b_{\lambda},b^\dagger_{\lambda'}\right]_-= \langle
\lambda|\lambda'\rangle\,;\label{eq9}
\end{equation}
$\lambda$ is the set of parameters (quantum numbers) specifying the
state; $E_\lambda=E_{-\lambda}>0$ is the energy of the state; symbol
$\sum\!\!\!\!\!\int\limits_{\lambda}$ denotes summation over
discrete and integration (with a certain measure) over continuous
values of $\lambda$; the ground state (vacuum) is defined
conventionally by relationship
\begin{equation}
        a_{\lambda}|{\rm vac}\rangle=b_{\lambda}|{\rm vac}\rangle=0.\label{eq12}
\end{equation}
Starting with the lagrangian for complex scalar field $\psi$ in the
form
$$
L=-\left[(\partial_\rho-{\rm i}\tilde{e}V_\rho)\psi\right]^*g^{\rho\rho'}
\left[(\partial_{\rho'}-{\rm i}\tilde{e}V_{\rho'})\psi\right]-(m^2+\xi R)\psi^*\psi,
$$
where $m$ is the mass of the scalar field and coupling $\tilde{e}$,
in general, differs from electromagnetic coupling $e$, and assuming
the ultrastaticity of the metric ($g_{00}=1$ and
$\partial_0g_{ii'}=0$) and the bundle ($V_0=0$ and
$\partial_0V_i=0$), we define functions $\psi_\lambda({\bf x})$
forming a complete set of solutions to stationary Klein--Gordon
equation
\begin{equation}
\fl\left[-g^{-1/2}(\partial_i-{\rm i}\tilde{e}V_i)g^{1/2}g^{ii'}(\partial_{i'}-
        {\rm i}\tilde{e}V_{i'})+m^2+\xi R\right]\psi_\lambda({\bf x})=E^2_\lambda\psi_\lambda({\bf x}),\label{eq11}
\end{equation}
and obeying orthonormalization condition
\begin{equation}
        \int d^dx\sqrt{g}\, \psi^*_\lambda({\bf x})\psi_{\lambda'}({\bf x})=\langle\lambda|\lambda'\rangle,\label{eq12}
\end{equation}
where $g={\rm det}g_{ii'}$.

In the case of space-time (5) with bundle connection (6) the
stationary Klein--Gordon equation outside the brane (at $r\neq 0$)
takes form
\begin{eqnarray}
\fl  \left[-r^{-1}\partial_rr\partial_r-(1-4G\mu)^{-2}r^{-2}\left(\partial_\varphi-
\frac{{\rm i}\tilde{e}\Phi}{2\pi}\right)^2-\Delta_{d-2}+m^2\right]
\psi_{kn{\bf p}}(r,\,\varphi,\,{\bf x}_{d-2})
=\nonumber \\ =({\bf p}^2+k^2+m^2)\psi_{kn{\bf
p}}(r,\,\varphi,\,{\bf x}_{d-2}),\label{eq13}
\end{eqnarray}
where $0<k<\infty$, $n\in\mathbb{Z}$, $-\infty<p^j<\infty$
($j=\overline{1,\,d-2}$). Imposing the Dirichlet boundary condition
on the brane\footnote{The most general boundary condition implies
that the scalar field is divergent but square integrable at $r=0$.
This condition involves at least four additional parameters
entailing a contact interaction with the brane (see \cite{Ada,Dab}
for the case of $d=2$ and $\mu=0$) and will be considered elsewhere.
Condition (14) corresponds to a particular choice of the parameter
values, when the contact interaction vanishes.},
\begin{equation}
   \psi_{kn{\bf p}}(0,\,\varphi,\,{\bf x}_{d-2})=0,\label{eq14}
\end{equation}
and normalizing in consistency with (12), one gets the solution to
(13):
\begin{equation}
    \psi_{kn{\bf p}}(r,\,\varphi,\,{\bf x}_{d-2})=\frac{(2\pi)^{(1-d)/2}}
    {\sqrt{1-4G\mu}}J_{|n-\tilde{e}\Phi/2\pi|(1-4G\mu)^{-1}}(kr)
  {\rm e}^{{\rm i}n\varphi}{\rm e}^{{\rm i}{\bf p}\cdot{{\bf x}_{d-2}}},\label{eq15}
\end{equation}
where $J_\omega(u)$ is the Bessel function of order $\omega$. As a
consequence of (14), there is no overlap between the region where
the matter is quantized ($r\neq 0$) and the region where the spatial
and the bundle curvatures are nonvanishing ($r=0$). In this sense,
the brane is impenetrable for quantum matter.

\section{Vacuum current}

The vacuum current of the scalar field is given by expression
\begin{equation}
  \fl {\bf j}({\bf x})=\frac{1}{4{\rm i}}\left\langle{\rm vac}\left|\left\{\left[\Psi^\dag(t,\,{\bf x}),\,
  \bnabla \Psi(t,\,{\bf x})\right]_+ -\left[\bnabla\Psi^\dag(t,\,{\bf x}),\,
  \Psi(t,\,{\bf x})\right]_+\right\}\right|{\rm vac}\right\rangle,\label{eq16}
\end{equation}
where symbol $[\ldots,\,\,\ldots]_+$ stands for the anticommutator,
and ${\bnabla}$ is the covariant derivative over spatial
coordinates. Using (8) and (15), we get $j_r={\bf j}_{d-2}=0$ and
\begin{eqnarray}
\fl  j_\varphi(r)\!=\!(2\pi)^{1-d}\!\int\! {\rm d}^{d-2}p\!\int\limits_{0}^{\infty}\!{\rm d}k\,k(p^2+k^2+m^2)^{-1/2}\times  \nonumber \\
  \times\sum\limits_{n\in\mathbb{Z}}(n-\tilde{e}\Phi/2\pi)(1-4G\mu)^{-1}J_{|n-\tilde{e}\Phi/2\pi|
  (1-4G\mu)^{-1}}^2(kr).\label{eq15}
\end{eqnarray}
As a manifestation of the Bohm--Aharonov effect, the vacuum current
is a periodic function of flux $\Phi$ with period equal to $2\pi
\tilde{e}^{-1}$, i.e. it depends on quantity
\begin{equation}
      F=\frac{\tilde{e}\Phi}{2\pi}-\left[\!\!\left[\frac{\tilde{e}\Phi}{2\pi}\right]\!\!\right],\label{eq16}
\end{equation}
where $\lshad u \rshad$ denotes the integer part of quantity $u$
(i.e. the integer which is less or equal to $u$). Note that relation
(18) can be rewritten as
\begin{equation}
F=\tilde{n}\tilde{e}/\tilde{e}_{\rm H}-\lshad\tilde{n}\tilde{e}/\tilde{e}_{\rm H}\rshad,\label{eq19}
\end{equation}
where $\tilde{n}\in \mathbb{Z}$ and $\tilde{e}_{\rm H}$ is the
coupling of the Higgs scalar field to bundle connection (6), since
the values of the flux are quantized as $\Phi=2\pi
\tilde{n}\tilde{e}_{\rm H}^{-1}$ \cite{Nie}.

Using relation
$$
J_\omega^2(u)=\frac{1}{{\rm i}\pi}[I_\omega(-{\rm i}u)K_\omega(-{\rm i}u)-I_\omega({\rm i}u)K_\omega({\rm i}u)]
$$
($I_\omega(u)$ and $K_\omega(u)$ are the modified Bessel functions
with exponential increase and decrease at large real positive values
of their argument), we get
\begin{eqnarray}
\fl j_\varphi(r)=\frac{1}{{\rm i}\pi}(2\pi)^{1-d}\int {\rm d}^{d-2}p\int\limits_{-\infty}^{\infty}{\rm d}k\,k(p^2+k^2+m^2)^{-1/2}
\times  \nonumber \\
\times \sum\limits_{n\in\mathbb{Z}}\nu(n-F)
I_{\nu|n-F|}(-{\rm i}kr)K_{\nu|n-F|}(-{\rm i}kr),\label{eq17}
\end{eqnarray}
\begin{figure}
  \includegraphics[width=260pt]{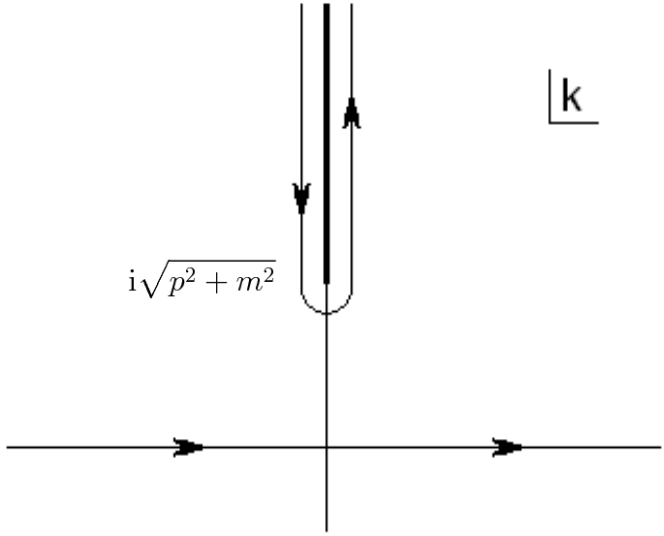}\\
  \caption{Branch point $k=\rmi\sqrt{p^2+m^2}$ and a cut in the upper half of the complex $k$-plane; the contour circumvents the cut.}\label{1}
\end{figure}\normalsize
where notation
\begin{equation}
\nu=(1-4G\mu)^{-1}\label{eq18}
\end{equation}
is used for brevity. The integral over real $k$ can be transformed
into the integral over a contour circumventing a part of the
imaginary axis in the complex $k$-plane, see figure 1. As a result,
we get
\begin{eqnarray}
\fl j_\varphi(r)\!=\!\frac{4}{(2\pi)^d}\!\int \!{\rm d}^{d-2}p
\!\int\limits_{\sqrt{p^2+m^2}}^{\infty}\!\!{\rm d}\kappa\,\kappa(\kappa^2\!-\!p^2\!-\!m^2)^{-1/2}
\sum\limits_{n\in\mathbb{Z}}\nu(n-F)
I_{\nu|n-F|}(\kappa r)K_{\nu|n-F|}(\kappa r)\!=  \nonumber \\
 =\frac{2}{(2\pi)^{d}}\int \!{\rm d}^{d-2}p
\!\int\limits_{\sqrt{p^2+m^2}}^{\infty}\!{\rm d}\kappa\,\kappa(\kappa^2-p^2-m^2)^{-1/2}
\!\int\limits_{0}^{\infty}\!\frac{{\rm d}y}{y}\,\exp\!\left(-\frac{\kappa^2r^2}{2y}-y\right)
\times \nonumber \\ \times\sum\limits_{n\in\mathbb{Z}}\nu(n-F)
I_{\nu|n-F|}(y),\label{eq19}
\end{eqnarray}
\begin{figure}
  \includegraphics[width=250pt]{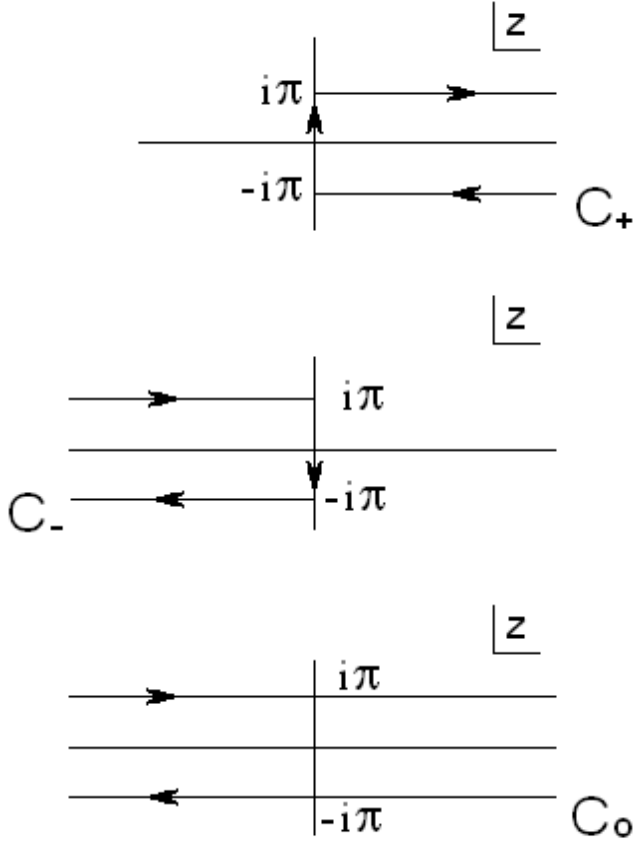}\\
  \caption{Contour $C_+$ or $C_-$ is used for the integral representation of $I_\omega (y)$ and contour $C_0$ is used for the integral representation of $K_\omega (y)$.}\label{2}
\end{figure}\normalsize
where in the last step the integral representation for the product
of modified Bessel functions is used (see, e.g., \cite{Pru}). Using
the Schl\"{a}fli contour integral representation
$$
I_\omega(y)=\frac{1}{2\pi {\rm i}}\int\limits_{C_+}{\rm d}z\,{\rm e}^{y\cosh z-\omega
z}=-\frac{1}{2\pi {\rm i}}\int\limits_{C_-}{\rm d}z\,{\rm e}^{y\cosh z+\omega z},
$$
we perform summation over $n$ and get
\begin{eqnarray}
\fl \sum\limits_{n\in\mathbb{Z}}\nu(n-F)I_{\nu|n-F|}(y)=\frac{y}{4\pi {\rm i}}
\int\limits_{C_0}{\rm d}z\,{\rm e}^{y\cosh z}\sinh z\frac{\sinh[(F-\frac{1}{2})\nu z]}{\sinh(\nu z/2)}=  \nonumber \\
=-\frac{y}{\pi}\int\limits_{0}^{\infty}{\rm d}u\,{\rm e}^{-y\cosh u}\sinh u
\,\Lambda\!\left(\cosh\frac u2;\,F,\,\nu\right),\label{eq20}
\end{eqnarray}
where contours $C_+$, $C_-$ and $C_0$ in the complex z-plane are
shown in figure 2, and
\begin{equation}
\fl \Lambda(v;\,F,\,\nu)=\frac{\sin(F\nu\pi)\sinh[2(1-F)\nu\,{\rm arccosh}\,v]-\sin[(1-F)\nu\pi]\sinh(2F\nu\,{\rm arccosh}\,v)}
  {\cosh(2\nu\,{\rm arccosh}\,v)-\cos(\nu\pi)}.\label{eq21}
\end{equation}
Substituting the last line of (23) into (22), integrating over
variables $\kappa$ and $y$, we get
$$
\fl j_\varphi(r)\!=\!-\frac{2}{(2\pi)^d}\sqrt{\frac{r}{\pi}}\int
\!{\rm d}^{d-2}p (p^2+m^2)^{3/4}\!\int\limits_{0}^{\infty}\!{\rm
d}u\,K_{3/2}\left(2r\sqrt{p^2+m^2}\,\cosh\frac u2\right)\times  $$
$$\times\frac{\sinh\frac u2}{\sqrt{\cosh\frac
u2}}\,\Lambda\!\left(\cosh \frac u2;\,F,\,\nu\right).
$$
Integrating over ${\bf p}$ and changing variable $u$ to
$v=\cosh\frac u2$, we get final expression for the vacuum current
\begin{equation}
\fl  j_\varphi(r)=-\frac{32}{(4\pi)^{(d+3)/2}}\frac{m^{(d+1)/2}}{r^{(d-3)/2}}
  \int\limits_{1}^{\infty}{\rm d}v\,v^{(1-d)/2}K_{(d+1)/2}(2mrv)\,\Lambda(v;\,F,\,\nu).\label{eq23}
\end{equation}

The vacuum current vanishes in the case of vanishing flux
($\Phi=0$), while in the case of vanishing tension ($\mu=0$) we get
\begin{equation}
  \Lambda(v;\,F,\,1)=-v^{-1}\sin(F\pi)\sinh[(2F-1){\rm arccosh}\,v],\label{eq24}
\end{equation}
and the result of \cite{Si9} is recovered
\begin{equation}
\fl  j_\varphi(r)=\frac{32\sin(F\pi)}{(4\pi)^{(d+3)/2}}\,\frac{m^{(d+1)/2}}{r^{(d-3)/2}}
  \!\int\limits_{1}^{\infty}\!\!{\rm d}v\,v^{-(d+1)/2}K_{(d+1)/2}(2mrv)\sinh[(2F-1){\rm arccosh}\,v], \, \nu=1.\label{eq25}
\end{equation}

Using the asymptotics of the Macdonald function at large values of
the argument,
$$
K_\omega(u)\,=\!\!\!\!\!\!\!\!\!_{_{u\rightarrow\infty}}\,{\rm e}^{-u}\sqrt{\frac{\pi}{2u}}[1+\Or(u^{-1})],
$$
we get the asymptotics of the vacuum current at large distances from
the brane:
\begin{equation}
\fl j_\varphi(r)\,=\!\!\!\!\!\!\!\!\!_{_{r\rightarrow\infty}}\frac{\nu\{F\sin[(1\!-\!F)\nu\pi]-(1\!-\!F)
\sin(F\nu\pi)\}}{(4\pi)^{(d+1)/2}\sin^2(\nu\pi/2)}{\rm e}^{-2mr}m^{(d-3)/2}r^{-(d+1)/2}\{1+\Or[(mr)^{-1}]\}.\label{eq26}
\end{equation}

\section{Vacuum magnetic field}

Assuming the valitity of the Maxwell equation
\begin{equation}
\frac{1}{(d-2)!}\nabla_{i_2}{\varepsilon^{i_1i_2}}_{i_3\ldots i_d}\,B_{({\rm I})}^{i_3\ldots i_d}=ej^{i_1},\label{eq27}
\end{equation}
one can deduce that, owing to the induced angular component of the
vacuum current, a magnetic field strength is also induced in the
vacuum outside the brane, being directed along it:
\begin{equation}
B^{3\ldots d}_{({\rm I})}(r)=\int\limits_{r}^{\infty}{\rm d}r\,\frac{\nu}{r}\,ej_\varphi(r);\label{eq28}
\end{equation}
note that coupling constant $e$ possesses dimension $m^{(3-d)/2}$ in
$(d+1)$-dimensional space-time. Substituting (25) into (30), we get
expression
\begin{equation}
\fl B^{3\ldots d}_{({\rm I})}(r)=-\frac{16e\nu}{(4\pi)^{(d+3)/2}}\left(\frac mr\right)^{(d-1)/2}
\int\limits_{1}^{\infty}{\rm d}v\,v^{-(d+1)/2}K_{(d-1)/2}(2mrv)\,\Lambda(v;\,F,\,\nu),\label{eq29}
\end{equation}
which decreases at large distances from the brane as
\begin{eqnarray}
\fl B_{({\rm I})}^{3\ldots d}(r)\,=\!\!\!\!\!\!\!\!\!_{_{r\rightarrow\infty}}\frac{e\nu^2\{F\sin[(1-F)\nu\pi]
-(1-F)\sin(F\nu\pi)\}}{2(4\pi)^{(d+1)/2}\sin^2(\nu\pi/2)}{\rm e}^{-2mr}m^{(d-5)/2}r^{-(d+3)/2}\times\nonumber \\
\times\{1+\Or[(mr)^{-1}]\}.\label{eq30}
\end{eqnarray}

Consequently, the magnetic flux is induced in the vacuum outside the
brane:
\begin{eqnarray}
\fl \Phi^{({\rm I})}=\int\limits_{0}^{2\pi}{\rm d}\varphi\int\limits_{0}^{\infty}{\rm d}
r\,\frac{r}{\nu}\,B_{({\rm I})}^{3\ldots d}(r)=
\nonumber \\ =-\frac{8em^{(d-1)/2}}{(4\pi)^{(d+1)/2}}\int\limits_{1}^{\infty}
{\rm d}v\,v^{-(d+1)/2}\Lambda(v;\,F,\,\nu)\int\limits_{0}^{\infty}{\rm d}r\,r^{(3-d)/2}K_{(d-1)/2}(2mrv).
\label{eq31}
\end{eqnarray}
In the case of ${\rm Re}\,d<3$ we get (see next section for details
and compare with relation (41))
\begin{equation}
\Phi^{({\rm I})}=\frac{e \,m^{d-3}\,\Gamma(\frac{3-d}{2})}{3(4\pi)^{(d-1)/2}}\left(F-\frac{1}{2}\right)F(1-F)\nu^2,\label{eq32}
\end{equation}
where $\Gamma(u)$ is the Euler gamma function; this result can be
analytically extended to a meromorphic function on the whole complex
$d$-plane, which in the physically meaningful domain ${\rm
Re}\,d\geq 2$ has poles at real odd values of $d$ and is finite at
real even values of $d$.

However, it seems more reasonable to restrict the validity of (34)
to the case of $d=2$,
\begin{equation}
\Phi^{({\rm I})}=\frac{e}{6m}\left(F-\frac 12\right)F(1-F)\nu^2,\qquad d=2,\label{eq33}
\end{equation}
while in the cases of $d=3,\,4,\ldots$ cutoff $r_0$ at small
distances to the brane should be introduced:
\begin{equation}
\Phi^{({\rm I})}=\frac{e}{6\pi}\left(F-\frac 12\right)F(1-F)\nu^2(-\ln mr_0),\qquad d=3,\label{eq34}
\end{equation}

\begin{equation}
\Phi^{({\rm I})}=-\frac{2e\,\Gamma\left(\frac{d-3}{2}\right)}{(4\pi)^{(d+1)/2}}
r_0^{3-d}\int\limits_{1}^{\infty}{\rm d}v\,v^{-d}\Lambda(v;\,F,\,\nu),\qquad d>3;\label{eq35}
\end{equation}
the integral in (37) can be explicitly taken in the case of real odd
values of $d$, see the next section.

\section{Massless quantum matter}

Using the asymptotics of the Macdonald function at small values of
the argument, we get the vacuum current for the massless scalar
field:
\begin{equation}
  j_\varphi(r)|_{m=0}=-\frac{16\,\Gamma\left(\frac{d+1}{2}\right)}{(4\pi)^{(d+3)/2}}r^{-d+1}
  \int\limits_{1}^{\infty}{\rm d}v\,v^{-d}\Lambda(v;\,F,\,\nu).\label{eq36}
\end{equation}
In the case of vanishing tension, the integral in (38) is taken,
yielding \cite{Si9}
\begin{equation}
\fl j_\varphi(r)|_{m=0}=\frac{4\sin(F\pi)}{(4\pi)^{(d+2)/2}}r^{-d+1}\left(F-\frac{1}{2}\right)
  \frac{\Gamma\left(\frac{d-1}{2}+F\right)\Gamma\left(\frac{d+1}{2}-F\right)}
  {\Gamma\left(\frac{d}{2}+1\right)}, \qquad \nu=1.\label{eq37}
\end{equation}
In the case of nonvanishing tension, the integral in (38) is
explicitly taken for odd values of $d$ only \cite{Dow}. To see this,
let us use the contour integral representation given by the first
line in (23) and present (38) in another way
\begin{equation}
\fl j_\varphi(r)|_{m=0}=\frac{\Gamma\left(\frac{d+1}{2}\right)}{(2\pi)^{(d+1)/2}}r^{-d+1}\frac{1}{4\pi {\rm i}}
  \int\limits_{C_0}{\rm d}z\frac{\sinh\,z}{(1-\cosh\,z)^{(d+1)/2}}\,\frac{\sinh\left[\left(F-\frac{1}{2}\right)\nu z\right]}
  {\sinh(\nu z/2)}.\label{eq38}
\end{equation}
If $d$ is odd, then the integrand in (40) has a pole at $z=0$ and
contour $C_0$ is deformed to encircle this pole, yielding
\begin{equation}
\fl j_\varphi(r)|_{m=0}=\frac{1}{6\pi^2}r^{-2}\left(F-\frac{1}{2}\right)F(1-F)\nu^2, \qquad d=3,\label{eq39}
\end{equation}
\begin{equation}
\fl j_\varphi(r)|_{m=0}=\frac{1}{12\pi^3}r^{-4}\left(F-\frac{1}{2}\right)F(1-F)\nu^2
  \left\{\frac{1}{3}+\frac{1}{5}\left[\frac{1}{3}+F(1-F)\right]\nu^2\right\}, \,\, d=5,\label{eq40}
\end{equation}
\begin{eqnarray}
\fl j_\varphi(r)|_{m=0}=\frac{1}{120\pi^4}r^{-6}\left(F-\frac{1}{2}\right)F(1\!-\!F)\nu^2\times \nonumber \\ \fl \times
\left\{\frac43+\left[\frac 13+F(1\!-\!F)\right]\nu^2+\frac17\left[\frac13+
F(1-F)+F^2(1\!-\!F)^2\right]\nu^4\right\}, \, d=7,\label{eq41}
\end{eqnarray}
\begin{eqnarray}
\fl j_\varphi(r)|_{m=0}\!=\!\frac{1}{560\pi^5}r^{-8}\!\left(\!F\!-\!\frac{1}{2}\!\right)F(1\!-\!F)\nu^2\times
\nonumber \\ \fl \times\left\{\!4+\frac{49}{15}\left[\frac13+F(1\!-\!F)\right]\nu^2+\frac23\left[\frac13+F(1\!-\!F)
+F^2(1\!-\!F)^2\right]\nu^4 +\right.\nonumber \\
\fl \left.+\frac{1}{15}\left[\frac13+F(1\!-\!F)+\frac{10}{9}F^2(1\!-\!F)^2+\frac
59 F^3(1\!-\!F)^3 \right]\nu^6\right\}, \, d=9,\label{eq42}
\end{eqnarray}
and so on; note that (41) was first obtained in \cite{Sri}. If $d$
is even, then the integrand in (40) has a branch point singularity
at $z=0$.

Using (30), we get expression for the vacuum magnetic field
strength:
\begin{equation}
\left.B_{({\rm I})}^{3\ldots d}(r)\right|_{m=0}=-\frac{8e\nu\Gamma\left(\frac{d-1}{2}\right)}{(4\pi)^{(d+3)/2}}r^{-d+1}\int\limits_{1}^{\infty}
{\rm d}v\,v^{-d}\Lambda(v;\,F,\,\nu).\label{eq43}
\end{equation}
Thus the induced vacuum flux is divergent either at small or at
large distances from the brane, or both. In the case of $d=2$,
introducing cutoff $r_\infty$ at large distances from the brane, we
get
\begin{equation}
\left.\Phi^{({\rm I})}\right|_{m=0}=-\frac{e}{2\pi}r_\infty\int\limits_{1}^{\infty}{\rm d}v\,v^{-2}\Lambda(v;\,F,\,\nu),\qquad d=2.\label{eq44}
\end{equation}
In the cases of $d=4,\,5,\,\ldots,$ introducing cutoff $r_0$ at
small distances to the brane, we get relation (37). In the case of
$d=3$ the vacuum flux is logarithmically divergent both at small and
at large distances from the brane, and, introducing both cutoffs, we
get
\begin{equation}
\left.\Phi^{({\rm I})}\right|_{m=0}=\frac{e}{6\pi}\left(F-\frac 12\right)F(1-F)\nu^2\ln\frac{r_\infty}{r_0},\qquad d=3.\label{eq45}
\end{equation}

\section{Discussion}

In the present paper we have obtained integral representation (25)
for the induced vacuum current of the quantized charged massive
scalar field in the cosmic string background. All the dependence on
the cosmic string parameters (tension $\mu$ and flux $\Phi$) is
described by function $\Lambda$ (24), whereas the dependence on
spatial dimension $d$ is factored otherwise. The Bohm-Aharonov
effect is manifested in the periodic dependence of the vacuum
current on flux $\Phi$ with period $2\pi\tilde{e}^{-1}$, i.e.
expression (25) depends on the fractional part of
$\tilde{e}\Phi/2\pi$ (18) rather than on quantity
$\tilde{e}\Phi/2\pi$ itself. Expression (25) vanishes at $F=1/2$
(i.e. at $\tilde{n}\tilde{e}=(n+1/2)\tilde{e}_{\rm H}$, where $n\in
\mathbb{Z}$ and $\tilde{n}\in \mathbb{Z}$), being negative and
convex downwards in the interval $0<F<1/2$, while positive and
convex upwards in the interval $1/2<F<1$; positions of the minimum
and the maximum are symmetric with respect to the point $F=1/2$.
Expression (25) can be analytically extended to complex values of
the space dimension yielding a holomorphic function of complex
variable $d$.

According to the Maxwell equation a magnetic field is generated by a
direct current, and thus the magnetic field strength is also induced
in the vacuum, being directed along the cosmic string, see (31). The
dependence of the field strength on the cosmic string parameters and
on the spatial dimension is qualitatively similar to that of the
current. Both the current and the field strength decrease
exponentially at large distances from the cosmic string, see (28)
and (32). Both the current and the field strength increase as
$r^{-d+1}$ in the vicinity of the cosmic string, see (38) and (45).
Thus the induced vacuum magnetic flux is finite in the case of $d=2$
only, see (35), whereas divergent otherwise, and a small vicinity of
the cosmic string is eliminated by cutoff $r_0$ in the cases of
$d>2$, see (36) and (37); note that the flux is hence independent of
mass $m$ in the cases of $d>3$. In the case of $d=3$, by identifying
the cutoff with the transverse size of the core of the cosmic
string, $r_0\simeq m_{\rm H}^{-1}$, we get
\begin{equation}
\Phi^{({\rm I})}=\frac{e}{6\pi}\left(F-\frac12\right)F(1-F)\nu^2\ln\frac{m_{\rm H}}{m},\qquad d=3.\label{eq46}
\end{equation}
One can conclude that, if the mass of the quantized scalar is
smaller than the mass of the Higgs scalar, $m\ll m_{\rm H}$, then
the cosmic string is enclosed in a tube of the magnetic flux lines
with the transverse size of the order of $(2m)^{-1}$.

It should be noted that the vacuum current and, consequently, the
vacuum magnetic field are odd under charge conjugation. The
production of cosmic magnetic fields during cosmological phase
transitions is currently discussed in the context of fundamental CP
violation in particle physics, see \cite{Vachas}. As it follows from
our consideration, topological defects (cosmic strings) appearing in
the aftermath of phase transitions can produce magnetic fields via
the vacuum polarization.

\ack{}

Yu\,A\,S acknowledges the support from the Department of Physics and
Astronomy of the National Academy of Sciences of Ukraine under\,
special program ``Fundamental properties of physical systems in
extremal conditions'' and from the State Foundation for Fundamental
Research of Ukraine under project F28.2/083 ``Application of the
string theory and field theory methods to the study of nonlinear
phenomena in low-dimensional systems''.

\end{document}